\documentclass[prb,a4paper,twocolumn,floatfix]{revtex4-1}
\usepackage{color}
\usepackage[usenames,dvipsnames]{xcolor}
\usepackage{graphicx}
\usepackage{latexsym}
\usepackage[]{amsmath}
\usepackage{epstopdf}
\usepackage{amssymb}
\usepackage{array}

\usepackage{dcolumn}
\usepackage{bm}

\begin{document}

\title{Fermi-surface topologies and low-temperature phases of the
filled Skutterudite compounds CeOs$_4$Sb$_{12}$ and NdOs$_4$Sb$_{12}$}

\author{Pei~Chun~Ho$^{1*}$, John~Singleton$^{2,3\aleph}$, 
Paul~A.~Goddard$^4$, Fedor~F.~Balakirev$^2$,
Shalinee~Chikara$^2$,
Tatsuya~Yanagisawa$^5$, M.~Brian~Maple$^{6,7}$,
David B.~Shrekenhamer$^{6\dag}$, Xia~Lee$^6$ and Avraham~T.~Thomas$^{6\ddag}$}

\affiliation{$^1$Department of Physics, California State University,
Fresno, CA~93740-8031, U.S.A.\\
$^2$National High Magnetic Field Laboratory, 
Los Alamos National Laboratory, MS-E536, Los Alamos, NM~87545, U.S.A.\\
$^3$University of Oxford, Department of Physics,
The Clarendon Laboratory, Parks Road, Oxford, OX1~3PU,
United Kingdom\\
$^4$Department of Physics, University of Warwick, Gibbet Hill Road, 
Coventry, CV4 7AL, United Kingdom\\
$^5$Department of Physics, Hokkaido University, Sapporo~060-0810, Japan\\
$^6$Department of Physics, University of California, San~Diego,
La~Jolla, CA~92093-0319, U.S.A.\\
$^7$Center for Advanced Nanoscience, University of California, San~Diego,
La~Jolla, CA~92093-0319, U.S.A. }

\begin{abstract}
MHz conductivity, torque magnetometer and magnetization
measurements are reported on single crystals of CeOs$_4$Sb$_{12}$ 
and NdOs$_4$Sb$_{12}$ using temperatures down to 0.5~K 
and magnetic fields of up to 60~tesla. 
The field-orientation dependence of the de Haas-van Alphen and
Shubnikov-de Haas oscillations is deduced
by rotating the samples about the $[010]$ and $[0\bar{1}1]$
directions. 
The results indicate that NdOs$_4$Sb$_{12}$ has
a similar Fermi surface topology to that of the unusual
superconductor
PrOs$_4$Sb$_{12}$, but with significantly smaller effective masses, 
supporting the importance of local phonon
modes in contributing to the low-temperature heat capacity of
NdOs$_4$Sb$_{12}$.
By contrast, CeOs$_4$Sb$_{12}$ undergoes a field-induced transition
from an unusual semimetal
into a high-field, high-temperature state 
characterized by a single, almost spherical
Fermi-surface section.
The behavior of the phase boundary and
comparisons with models of the bandstructure lead us to
propose that the field-induced phase transition in
CeOs$_4$Sb$_{12}$ is similar in origin to the
well-known $\alpha-\gamma$ transition in Ce and
its alloys.
\end{abstract}


\maketitle

\section{Introduction}
Filled skutterudite compounds, with the formula
MT$_4$X$_{12}$, where  M is
an alkali metal, alkaline-earth, lanthanide, or actinide,
T is Fe, Ru, or Os and X is P, As, or Sb, display a wide
variety of interesting phenomena caused by
strong electron 
correlations~\cite{tayama2015,yan,HoNJP,aokireview,oneprime,
meisner,Sugawara1,Sugawara,Sugawara3}.
Amongst these, the three compounds   
CeOs$_4$Sb$_{12}$, PrOs$_4$Sb$_{12}$, and NdOs$_4$Sb$_{12}$,
formed by employing Periodic-Table neighbors for M,
span the range from an antiferromagnetic (AFM) 
semimetal~\cite{tayama2015,yan} or perhaps
Kondo insulator (M~=~Ce)~[\onlinecite{BauerED2001}] 
via a 1.85~K unconventional (quadrupolar-fluctuation mediated~\cite{Miyake2003})
superconductor (M~=~Pr)~[\onlinecite{Sugawara}] 
to a 1~K ferromagnet (FM; M~=~Nd)~[\onlinecite{Ho2005}].
In view of the low ordering temperatures, all around $1-2$~K, 
associated with their various groundstates, plus precedents
in other unconventional superconductors~\cite{heavyfermion,merino}, 
it is tempting to
speculate as to whether the superconductivity in M~=~Pr arises
from close proximity to AFM and FM quantum-critical points,
with the more explicitly magnetic
M~=~Ce and M~=~Nd lying on the either side of the postulated
quantum phase transitions.
To explore this idea further, knowledge of the 
Fermi surfaces of all three materials is essential.
Whilst the Fermi surface of PrOs$_4$Sb$_{12}$ is relatively well
known~\cite{Sugawara}, those of 
CeOs$_4$Sb$_{12}$ and NdOs$_4$Sb$_{12}$
have been little studied.
In this paper, we therefore use magnetic fields of up to 60~T
to measure Fermi-surface cross-sections
and effective masses for these two
skutterudites. In the process we find a new 
magnetic-field-driven phase transition in CeOS$_4$Sb$_{12}$,
whilst our measurements of NdOs$_4$Sb$_{12}$ point to the
contribution of local phonon modes to the low-temperature
heat capacity.

The paper is organized as follows. Experimental details, including crystal
growth and measurement technques,
are given in Section II. Sections III discusses
high-field data from CeOs$_4$Sb$_{12}$, 
including evidence for the field-induced 
change in Fermi surface
and the delineation of
the new phase boundary at which this occurs, 
whilst the Fermi Surface of NdOs$_4$Sb$_{12}$
is treated in Section IV.
Section V presents a summary and conclusions.
 
\section{Experimental details}
\label{experiment}
\subsection{Crystal growth}
\begin{figure}[t]
\centering
\includegraphics[width=6.5cm]{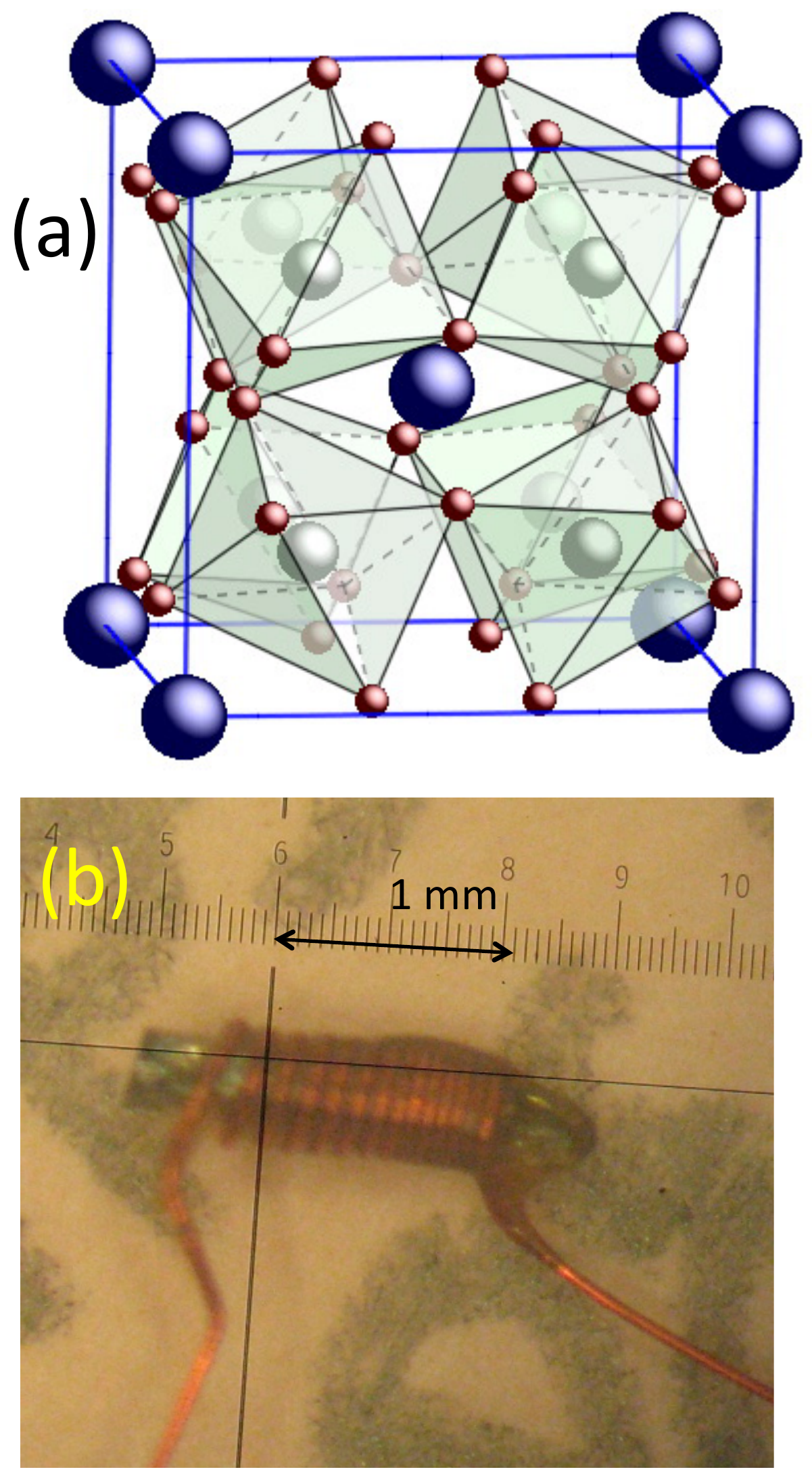}
\sloppypar
\vspace{-3mm}
\caption{(a)~Cubic (non-primitive) unit cell for 
NdOs$_4$Sb$_{12}$ and CeOs$_4$Sb$_{12}$.
Nd/Ce (corners and body center position) are purple;
Os ions are gray, surrounded by cages with red
Sb ions at the vertices.
(b)~Typical NdOs$_4$Sb$_{12}$ single
crystal with proximity-detector oscillator
coil (insulated 46 gauge, high-conductivity copper)
wrapped around it.
}
\vspace{-5mm}
\label{fig1}
\end{figure}

Single crystals of CeOs$_4$Sb$_{12}$ and NdOs$_4$Sb$_{12}$
are synthesized using a molten-flux technique~\cite{BauerED2001} 
with an excess of Sb (40 to 1 ratio relative to the rare-earth element). 
The purities of the rare-earth elements are 3N; those of
Os and Sb 4N and 5N respectively.  
Typical dimensions of the single crystals
used in the experiments in this paper
 are $\sim 3.5 \times 0.4 \times 0.3~{\rm mm}^3$ for
both compounds.
The crystals have cubic space group Im$\bar{3}$; the
cubic (non-primitive) unit cell and ion positions are shown in Figure~\ref{fig1}(a).
Further structural details are given in Refs.~\onlinecite{maplepnas}
and~\onlinecite{yuhaszprb}.

\subsection{Quantum oscillation measurements}
Both de Haas-van Alphen and Shubnikov-de Haas
oscillations are used to determine the Fermi-surface
cross-sections of the two materials
as a function of the orientation of the
applied magnetic field. 
Oscillation frequencies and effective
masses deduced from both techniques are found
to be in good agreement.
In addition, data taken using different crystals
and batches of the same material are in accord
within experimental errors,
suggesting that the features described below are intrinsic
to the two compounds.

Measurements of the de Haas-van Alphen effect
are carried out in quasistatic magnetic fields
and  employ
a torque magnetometer with a cantilever constructed from $5~\mu$m
phosphor bronze~\cite{HoNJP}. 
A small slice taken from a single crystal is glued
to the cantilever via a thin sheet
of strain-reducing paper. The interaction
between the sample's magnetic moment {\bf m}
and the applied magnetic field {\bf B}
causes a torque $-{\bf m} \times {\bf B}$
that results in a deflection of the cantilever.
The deflection is monitored using the capacitance
between the cantilever and a fixed plate about 1~mm below it
($\sim 0.5$~pF), measured using a General Radio Capacitance bridge.
Care is taken to ensure that deflections are small,
so that the sample's orientation in the field is
not changed significantly by the torque.
The torque magnetometer is mounted on a two-axis
cryogenic goniometer that allows the sample orientation
to be changed {\it in situ}; $^3$He refrigeration
provides temperatures in the range $0.45-10$~K.
Quasistatic magnetic fields are provided by
a 35~T Bitter coil at NHMFL Tallahassee.

Shubnikov-de Haas oscillations are measured in pulsed 
magnetic fields using the contactless-conductivity 
method described in Refs.~\onlinecite{saman} and \onlinecite{moaz}.
A coil comprising $5-12$ turns of 46-gauge
high-conductivity copper wire is wound about the
single-crystal sample (see Figure~\ref{fig1}(b)); the number of 
turns employed depends on the
cross-sectional area of the sample, with a larger number of turns
being necessary for smaller samples.
The coil forms part of a proximity detector oscillator (PDO) circuit 
resonating at 22-29~MHz. 
A change in the sample skin depth~\cite{moaz} or 
differential susceptibility~\cite{saman}
causes a change in the inductance 
of the coil, which in turn alters the resonant frequency of the circuit. 
Shubnikov-de Haas oscillations are observed in the resistivity 
and hence the skin depth and frequency~\cite{saman,moaz}.
For the purposes of digitizing the data
prior to Fourier transformation to obtain the frequency
as a function of field,
the signal from the PDO circuit is
mixed down to about 2~MHz using a double heterodyne
system~\cite{saman,moaz}.
Data are recorded at 20~Msamples/s,
well above the Nyquist limit.
Two or three samples in individual coils coupled to
independent PDOs are measured 
simultaneously, using a single-axis, worm-driven,
cryogenic goniometer to adjust their orientation in the field.
Pulsed magnetic fields are provided by the
60~T long-pulse magnet
and one of the 65~T short-pulse magnets at NHMFL Los Alamos~\cite{nhmfl};
the former magnet, with its relatively slow
rise time, is used to check that
inductive sample heating is not an issue.

The goniometer is placed within a
simple $^3$He cryostat providing temperatures
down to 0.4~K; temperatures are measured using
a Cernox sensor supplied and calibrated by Lakeshore Inc.
Magnetic fields are deduced by integrating the
voltage induced in an eleven-turn coil,
calibrated against the de Haas-van Alphen
oscillations of the belly orbits of copper~\cite{HoNJP}.
In both experiments, the crystals were either
rotated about their $[010]$ axis or their $[1\bar{1} 0]$ axis,
with the axis of rotation lying perpendicular to the magnetic field. 

The phase boundary shown in Section~III was traced using
both the pulsed-field extraction magnetometer described in Ref.~\onlinecite{Goddard}
and a commercial SQUID magnetometer (Quantum Design MPMS with 7~T magnet).

\section{Results on Cerium Osmium Antimonide}
\subsection{Overview of CeOs$_4$Sb$_{12}$ PDO data; phase boundary and quantum oscillations}
\begin{figure}[t]
\centering
\includegraphics[width=8.5cm]{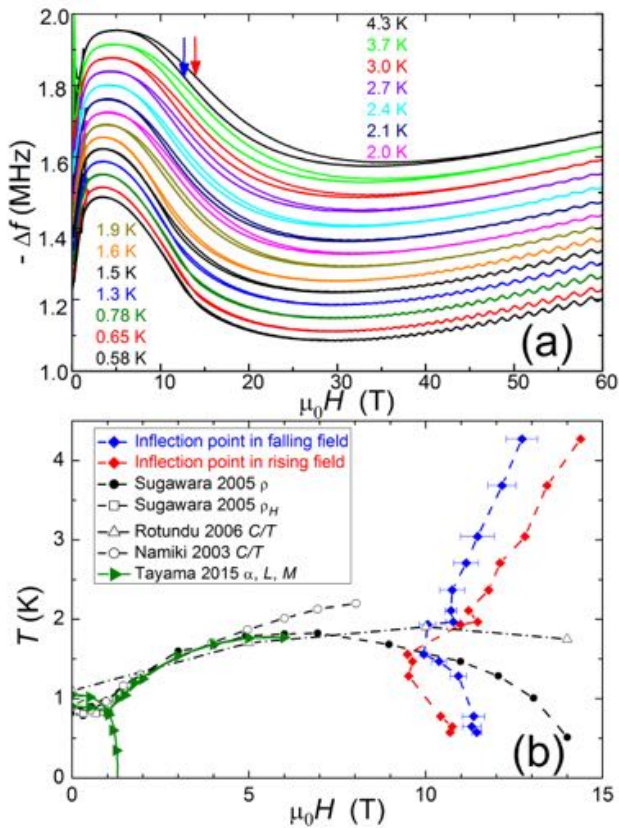}
\sloppypar
\caption{(a)~The negative of the PDO frequency shift,
$-\Delta f$, observed for a CeOs$_4$Sb$_{12}$ single crystal
plotted as a function of magnetic field for
several sample temperatures $T$; the color key for the
various $T$ is given within the figure.
{\bf H} is parallel to [100].
Data for both up- and down-sweeps of the
field are shown; the blue (red) arrow
indicates the position of the inflection
point of the $-\Delta f$ versus $H$ curve
in falling (rising) field for the 4.3~K data set.
Note Shubnikov-de Haas oscillations emerging
at higher fields.
(b)~Inflection points in the $-\Delta f$ versus $H$
curves shown in (a) plotted as a function of $T$.
Notional phase boundaries supposedly enclosing 
the low-temperature antiferromagnetic state and
deduced by other authors (Namiki {\it et al.}~[\onlinecite{Namiki2003}], 
Sugawara {\it et al.}~[\onlinecite{Sugawara2005}],
Rotundu {\it et al.}~[\onlinecite{Rotundu}],
Tayama {\it et al.}~[\onlinecite{tayama2015}]) are shown
for comparison; techniques and authors are shown in the
inset key.}
\label{CeSweep}
\end{figure} 
Figure~\ref{CeSweep} shows the PDO frequency as a function of
field for a variety of temperatures. 
As mentioned above, an increase in sample skin depth,
due to an increase in resistivity, will raise the inductance
of the coil, as the sample excludes flux from a smaller area.
Hence the PDO resonant frequency $f$ will drop.
For small changes it can be shown that~\cite{saman}
\begin{equation}
\Delta f \propto -\Delta \rho
\end{equation}
where $\Delta f$ is the shift in frequency due to a 
change in sample resistivity $\Delta \rho$.
Hence, Figure~\ref{CeSweep} has been plotted as
$-\Delta f$ versus $H$,
so that upward shifts in the data indicate increasing
resistivity.

As the field increases, the resistivity rises steeply to a broad
peak before falling to a minimum at about 25~T;
note that within the region where the resistance falls
there is hysteresis between the up- and downsweeps
of the magnetic field, suggestive of a phase boundary~\cite{fivos}.
From now on, we shall refer to the field regions
on either side of this boundary as the L (low-field, low-temperature) 
and H (high-field, high-temperature) phases.

Following precedents set by similar, broad phase transitions~\cite{fivos},
we plot the inflection point in the $-\Delta f$ versus $H$
curves (see arrows in Fig.~\ref{CeSweep}(a)); 
in Fig.~\ref{CeSweep}(b) in red (rising field) and blue (falling field). 
In previous work on CeOs$_4$Sb$_{12}$, 
a phase transition at $T\approx 1$~K
with low entropy was detected; results from magnetoresistance, 
Hall effect, Sb nuclear quadrupole resonance (NQR),
dilatometry 
and neutron-scattering experiments indicate that this 
low-temperature phase is intrinsic and originates from an 
antiferromagnetic (AFM) state, perhaps associated with a 
spin-density-wave (SDW) ground 
state.~\cite{tayama2015,Sugawara2005,Yogi2005,YangCP2006,YangCP2007}  
Notional phase boundaries enclosing 
this AFM state,
deduced by other authors~\cite{tayama2015,Namiki2003,Sugawara2005,Rotundu} are shown
in Figure~\ref{CeSweep}(b) for comparison.
The L to H transition marked by the fall in $-\Delta f$ around 11~T
(red and blue points in Figure~\ref{CeSweep}(b)) 
persists to temperatures well above the phase boundary 
around the antiferromagnetic state (black and hollow points).
This shows that, above 2~K, the L to H transition
cannot be {\it caused} by
a field-induced destruction of the antiferromagnetic state ({\it c.f.}
Ref.~\onlinecite{Harrison}).

However, below 1.5~K and above 10~T, 
we believe that the L to H transition
coincides with the boundary of the
antiferromagnetic phase.
In Section IIIC, we discuss changes
in Fermi-surface topology and effective Ce valence
that occur at the L to H transition;
these are almost certain to destroy any antiferromagnetism.
In this context,
the features in the original data~\cite{Sugawara2005,Rotundu} 
used by others to denote the high-field limit
of the antiferromagnetic phase
(Fig.~\ref{CeSweep}(b), hollow triangles, black, filled circles)
are rather broad.
Given that the fall in $-\Delta f$ covers several tesla
(Fig.~\ref{CeSweep}(a)), we believe that the
points measured below 1.5~K and between 10 and 15~T in
Refs.~\onlinecite{Sugawara2005,Rotundu}
are in fact associated with the broadened L to H transition
that we observe in the PDO data.
Both of these collections 
of points lie within the extended field range
of the fall in $-\Delta f$, and the scatter between 
the two data sets probably reflects the
difficulty of assigning a precise field to
such a broadened feature~\cite{fivos}.
The switching of the order of the up- and downsweep
L to H PDO inflection points at low temperatures (Fig.~\ref{CeSweep}(b),
red and blue points)
may well be a manifestation of the 
interdependence of the antiferromagnetic
and L phases, a point to which we return in Section IIID. 

At higher fields, (in the H phase),
Shubnikov-de Haas oscillations can be seen
superimposed on a gentle positive magnetoresistance
(see Figures~\ref{CeSweep}(a) and \ref{CeFreqs1}(a)).
We shall now discuss these quantum oscillations,
returning later to the origin of the
L to H transition.

\subsection{Shubnikov-de Haas oscillations and Fermi-surface topology in the H phase}
\begin{figure}[t]
\centering
\includegraphics[width=8.5cm]{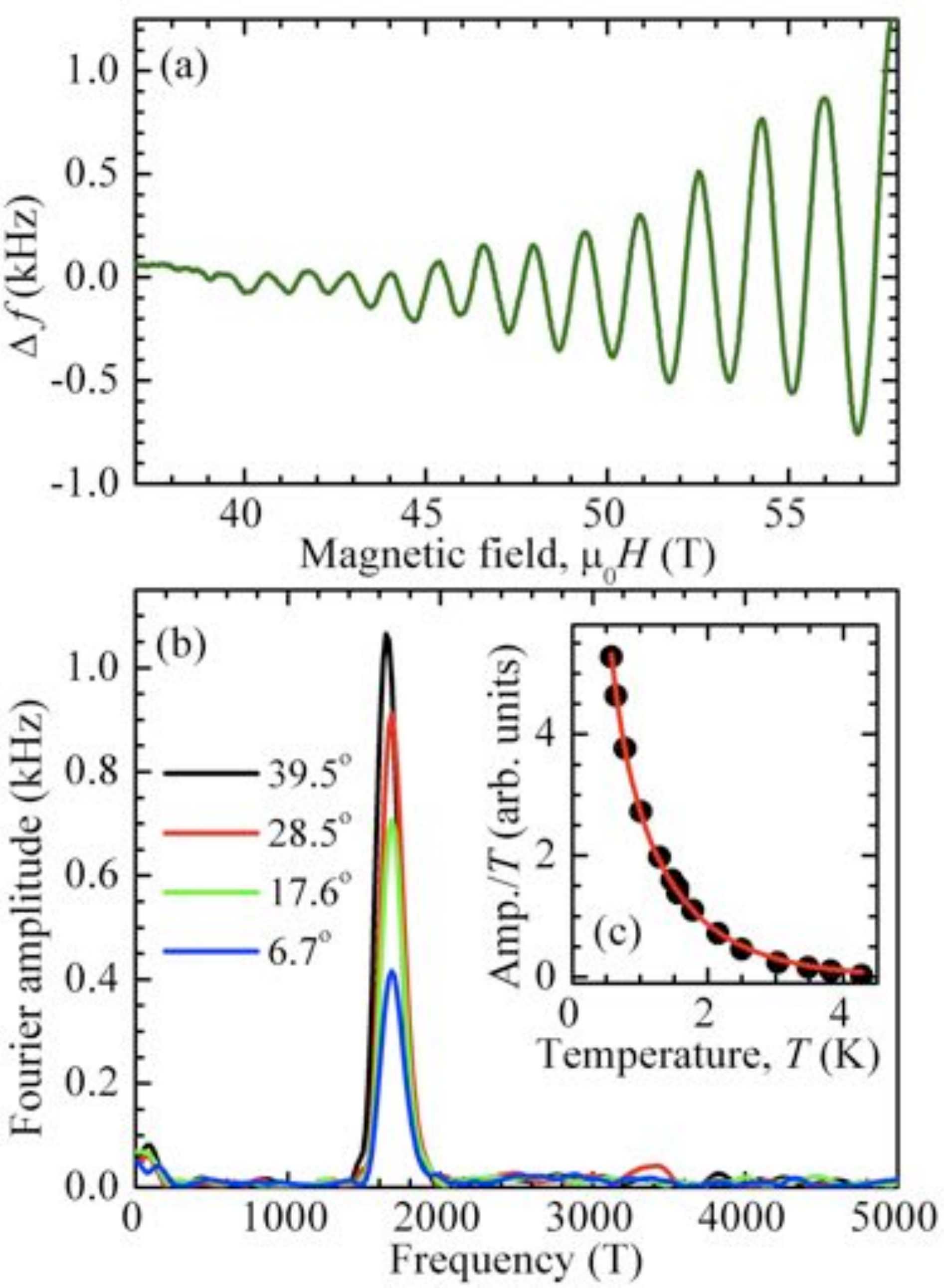}
\sloppypar
\caption{(a)~Shubnikov-de Haas oscillations 
in the PDO frequency shift of a CeOs$_4$Sb$_{12}$
single crystal,
obtained by subtracting the monotonic 
background dependence due to
magnetoresistance. The temperature is
$T=1.5$~K and the field is oriented at
$28.5^{\circ}$ to the $[0\bar{1}0]$ axis
for rotation about the $[110]$ axis.
(b) Fourier transforms of data similar to those
in (a) for the orientations shown in the inset key
with respect to the $[0\bar{1}0]$ axis;
$T = 1.5$~K and the sample is
rotated about the $[110]$ axis.
(c)~CeOs$_4$Sb$_{12}$
Shubnikov-de Haas-oscillation
Fourier amplitude divided by $T$ versus $T$;
data are points and the curve is a fit to
Equation~\ref{LK}.
{\bf H} was applied parallel to the [001] direction.}
\label{CeFreqs1}
\end{figure}

A single frequency of Shubnikov-de Haas oscillations
is observed in the H phase at all angles of the field;
data such as those in Figure~\ref{CeFreqs1}(b)) show that
there is comparitively little angular variation of the
extremal orbits, suggesting that the Fermi surface
is likely to be an approximate sphere centred on the
$\Gamma$ point. 
The quasiparticle effective mass $m^*$ was evaluated by examining 
the temperature ($T$) dependence of the Shubnikov-de Haas
oscillation amplitude $A$ (Figure~\ref{CeFreqs1}(c)),
and fitting it to the following portion of the
Lifshitz-Kosevich formula~\cite{shoenberg}:
\begin{equation}
\frac{A}{T} \propto \frac{1}{B \sinh (\frac{14.7T}{B}\frac{m^*}{m_{\rm e}})};
\label{LK}
\end{equation}
here, $m_{\rm e}$ is the mass of a free electron.
A relatively light value of $m^*/m_{\rm e} =  3.6\pm 0.1$ was obtained for
${\bf H}|| [010]$.

The full angle dependence of the Shubnikov-de Haas
oscillation frequency (and, via the Onsager relationship~\cite{shoenberg},
the Fermi-surface cross-section)
of CeSb$_4$Os$_{12}$ in the high-field phase
is shown in Figure~\ref{CeArea}(a);
for comparision, data for CeRu$_4$Sb$_{12}$
from Ref.~\onlinecite{Sugawara3} are also shown.

The low-temperature electrical resistivity of 
CeOs$_4$Sb$_{12}$ was initially interpreted in terms of  
a Kondo insulator with a small energy 
gap of $\approx 10$~K~[\onlinecite{BauerED2001}].
However, bandstructure calculations~\cite{yan,HarimaPC} 
suggest that CeOs$_4$Sb$_{12}$ is a compensated semimetal,
with the chemical potential lying close to the bottom of 
an electron-like band (with very high density of states)
and the top of a hole-like band,
producing a Fermi surface consisting of tiny
hole and electron pockets;
this is shown schematically in Fig.~\ref{CeArea}(b).
Yan {\it et al.}~[\onlinecite{yan}] remark
that CeOs$_4$Sb$_{12}$ is very close
to being a topological insulator~\cite{yan}.
In the absence of symmetry breaking (for example
caused by strain), the energy gap is zero, leading to
the above-mentioned Fermi-surface pockets~\cite{yan,HarimaPC}.
The predicted Shubnikov-de Haas frequencies that would result
from these calculated Fermi surfaces are much smaller than
the experimental values shown in Fig.~\ref{CeArea}(a).

In Ref.~\onlinecite{yan},
Yan {\it et al.} calculate the effect of removing electrons
from the CeOs$_4$Sb$_{12}$ bandstructure;
the result is a single, much larger, approximately
spherical hole pocket centred on the
$\Gamma$ point (shown schematically in Fig.~\ref{CeArea}(c)).
(The same conclusion 
can be derived from Fig.~3
of Ref.~\onlinecite{HarimaPC}.)
Such a Fermi-surface toplogy is qualitatively
similar to our experimental observations in the
H phase of CeOs$_4$Sb$_{12}$, providing,
as we shall see in the following Section, 
a clue as to the cause of the
L to H phase transition observed in Fig.~\ref{CeSweep}.

\subsection{Origin of the L to H phase boundary in CeOs$_4$Sb$_{12}$}
Based on experimental evidence, we 
propose that the L to H phase transition
in CeOs$_4$Sb$_{12}$ is a
{\it valence transition}~\cite{fivos,dzero} analogous to those
observed in YbInCu$_4$~[\onlinecite{dzero,Sarrao1,Sarrao2,Sarrao3}], 
elemental Ce~[\onlinecite{dzero,Fi02,Fi05,Fi06}] and Ce alloys 
such as Ce$_{0.8}$La$_{0.1}$Th$_{0.1}$~[\onlinecite{fivos}].
In these materials, the valence transition occurs between
a higher-temperature, higher-field phase in which 
quasi-localized 4$f$ moments (on the Yb or Ce ions) 
are stabilized by entropy terms in the
free energy, and a band-like
state of the $f$ electron at low temperatures and
low fields~\cite{dzero}. For brevity,
we refer to these phases as H and L respectively,
just as as in the case of CeOs$_4$Sb$_{12}$.

\begin{figure}[t]
\centering
\includegraphics[width=8.5cm]{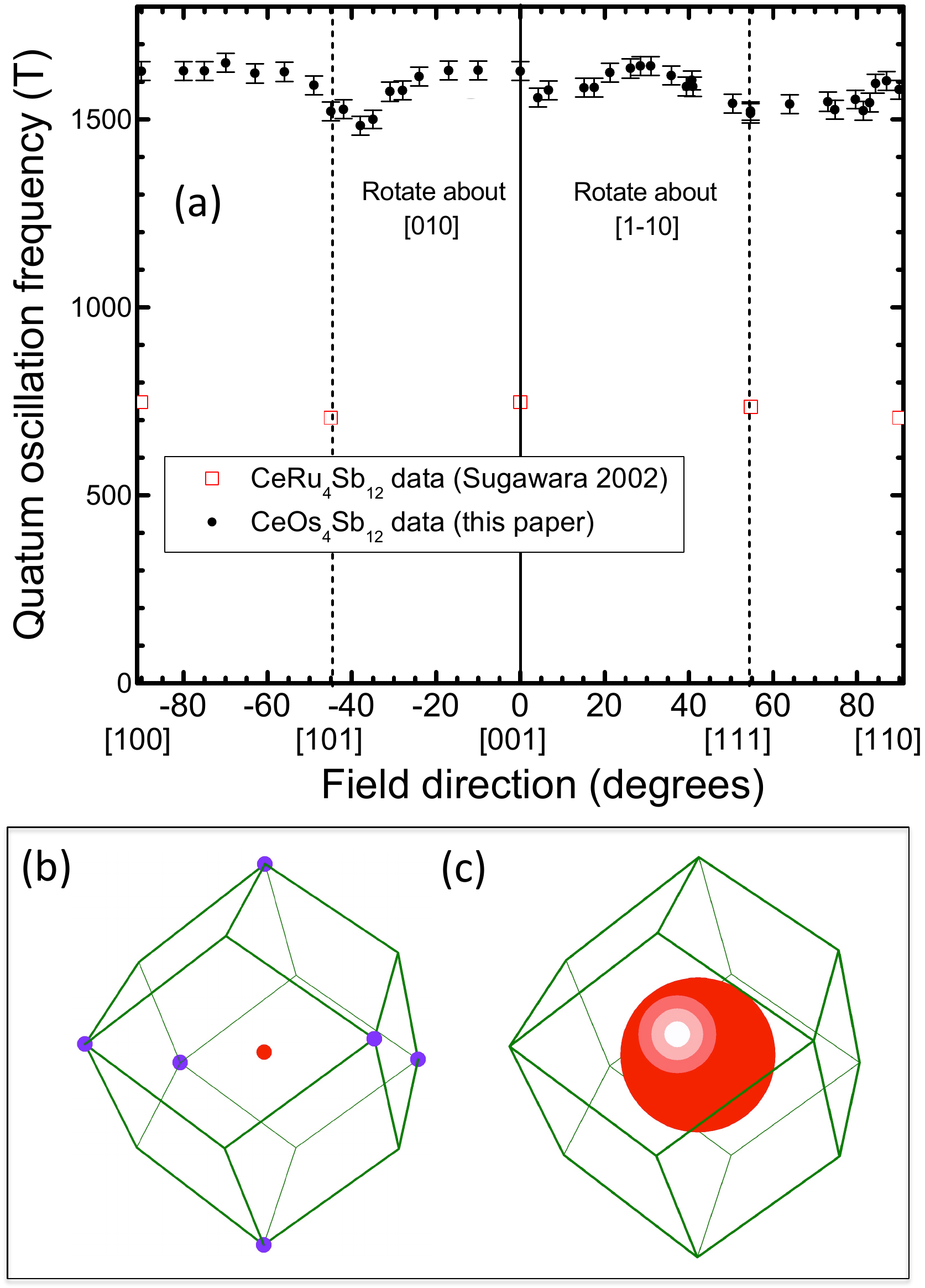}
\sloppypar
\caption{(a)~Field-orientation dependence of the
Shubnikov-de Haas oscillation frequency
of the H phase of CeOs$_4$Sb$_{12}$
(black points);
For comparison, analogous data from
CeRu$_4$Sb$_{12}$ taken from Ref.~\onlinecite{Sugawara3}
are shown in red.
(b)~Schematic of semimetallic Fermi 
surface for CeOs$_4$Sb$_{12}$
and first Brillouin zone (dark green lines)
based on Ref.~\onlinecite{yan}.
A tiny hole pocket at the zone-center $\Gamma$ point
(red) is compensated by electron pockets
(purple) at the H points of the Brillouin zone;
$\frac{1}{6}$ of each electron pocket is in the
first Brillouin zone.
(c)~Schematic of CeOs$_4$Sb$_{12}$ Fermi surface
after a downshift in chemical potential
corresponding to the removal
of itinerant quasiparticles from the bandstructure
(based on Refs.~\onlinecite{yan,HarimaPC}).
The discussion in Section~IIIC suggests that
the Fermi surface in the L phase is
similar to that shown in (b), whilst the H-phase
Fermi surface is similar to that shown in (c).
}
\label{CeArea}
\end{figure}  

In the above materials,
both the L and H phases are thought to possess
correlated 4$f$  electrons, but with very different
effective Kondo temperatures~\cite{fivos,dzero}.
The effective Kondo temperatures 
of the H phases are small, so that
the properties of the 4$f$ electrons will 
be almost indistinguishable from those of 
localized ionic moments~\cite{fivos,dzero}; any remaining 
itinerant quasiparticles will have relatively
light effective masses (as we observe experimentally
in CeOs$_4$Sb$_{12}$; see the discussion of Fig.~\ref{CeFreqs1}(c)).
By contrast, the L phases 
of the above-mentioned systems have relatively
large Kondo temperatures. This will
cause the 4$f$  electrons to be in the mixed-valence regime
with signifcant $spd$  and $f$  hybridization,
resulting in itinerant quasiparticles with
a large effective mass~\cite{fivos,dzero,Sarrao1}.
This is again in accord with the behavior of CeOs$_4$Sb$_{12}$;
the L phase has an estimated Kondo temperature 
$T_{\rm KL}\approx 100$~K~[\onlinecite{BauerED2001}],
and evidence for a very heavy effective mass is
given by heat-capacity data~\cite{BauerED2001,Namiki2003,Rotundu} 
that reveal a relatively
large electronic heat-capacity coefficient
$\gamma= 92 - 180~{\rm mJmol^{-1}K^{-2}}$,
despite the small itinerant carrier density~\cite{HarimaPC}.

In all such systems, the L-phase
itinerant quasiparticles will be preferable 
to quasi-localized electrons from
an energetic standpoint 
–- they have smaller zero-point energy -- at low temperatures,
but quasi-localized 4$f$ electrons will be 
favoured on entropic grounds at elevated temperatures~\cite{dzero}.
Therefore, increasing the temperature will drive the L to H transition.
Similarly, a magnetic field affects
the free energy of the quasi-localized 4$f$ electrons
in the H phase more strongly than the energies of 
the band quasiparticles
in either the L or H phases~\cite{fivos,dzero}.
Thus, the H phase becomes more energetically
favorable with increasing field, so that
a field-driven L to H transition occurs,
as observed in CeOs$_4$Sb$_{12}$
and the other systems~\cite{fivos,Sarrao1,Sarrao2,Sarrao3}.

In common with our observations of CeOs$_4$Sb$_{12}$,
the valence transitions from L to H 
phases in YbInCu$_4$, Ce and Ce alloys
are marked by (very frequently broadened) 
changes in resistivity and
magnetic susceptibility, both accompanied
by hysteresis in field and temperature
sweeps~\cite{fivos,Sarrao1,Sarrao2,Sarrao3,Fi02,Fi05,Fi06}.
This type of transition is most spectacular in
elemental Cerium and Ce alloys~\cite{fivos,Fi02,Fi05,Fi06}, where
the L phase is known as the $\alpha$ phase, and H as 
the $\gamma$ phase~\cite{fivos,Fi02,Fi05,Fi06}.
Whereas the 4$f$ ions in CeOs$_4$Sb$_{12}$ and
YbInCu$_4$ represent a relatively small proportion
of the elements in their 
respective solids, in Ce itself, {\it every}
ion is involved, and the $\gamma-\alpha$ transition is accompanied
by a 14.8\% volume collapse as the redistribution of charge
between ions and the sea of itinerant quasiparticles occurs;
nevertheless, the face-centred-cubic (fcc)
structure of Ce is preserved~\cite{fivos,beta}.
This fascinating aspect of Cerium's behavior
means that the $\gamma-\alpha$ transition
has attracted considerable attention, especially 
in the metallurgy community (see Ref.~\onlinecite{Fi02}
and references therein).

In the L phase of CeOs$_4$Sb$_{12}$,
all of the available electrons will contribute
to the sea of itinerant quasiparticles;
CeOs$_4$Sb$_{12}$ will be a compensated 
semimetal~\cite{yan,HarimaPC}
with a Fermi surface similar to that shown schematically in
Fig.~\ref{CeArea}(b).
Though the carrier density is small,
bandstructure calculations~\cite{HarimaPC}
suggest that the chemical potential is located in
a region where there exists
a very large density of states associated with 
bands derived mostly from
the 4$f$ orbitals, leading to a large
electronic heat capacity~\cite{BauerED2001,Namiki2003,Rotundu}. 
On moving from the L to the H phase,
the removal of $f$ electrons from
the itinerant quasiparticle sea will
have the effect of lowering the chemical potential, 
leading to a hole Fermi surface at the $\Gamma$ point,
similar to that shown in Fig.~\ref{CeArea}(c)~[\onlinecite{yan}].
In CeOs$_4$Sb$_{12}$,
the transition from a small density of very heavy quasiparticles (L phase)
to a larger Fermi surface of much lighter holes (H phase) should lower the 
resistivity, as is indeed observed experimentally (Fig.~\ref{CeSweep}(a));
this is in contrast to the case of Ce, where the very different bandstructure
leads to a rise in resistance on going from $\alpha$ (L) to
$\gamma$ (H)~[\onlinecite{fivos}].
Finally, as the chemical potential in CeOs$_4$Sb$_{12}$
moves down at the L to H transition, away from
the bands with the high density of states~\cite{HarimaPC},
the electronic contribution to the heat capacity should also fall,
despite the larger Fermi surface~\cite{HarimaPC,HarimaAPS};
experiments carried out at fields close to the L to H border
seem to suggest that 
the electronic heat-capacity coefficient $\gamma$ indeed falls with 
increasing field~\cite{Rotundu}.

The antiferromagnetic state of CeOs$_4$Sb$_{12}$
has been attributed to a possible SDW
state~\cite{tayama2015,Sugawara2005,Yogi2005,YangCP2006,YangCP2007}.
SDWs are usually associated with features of the
Fermi-surface topology~\cite{mybook};
therefore, as the Fermi surface changes
as one crosses from L to the H phase, one would expect the
antiferromagnetic state to be destroyed,
as we argued in the discussion of 
Fig.~\ref{CeSweep} above.
\subsection{Delineating the L to H phase boundary in CeOs$_4$Sb$_{12}$}
\begin{figure}[t]
\centering
\includegraphics[width=8.5cm]{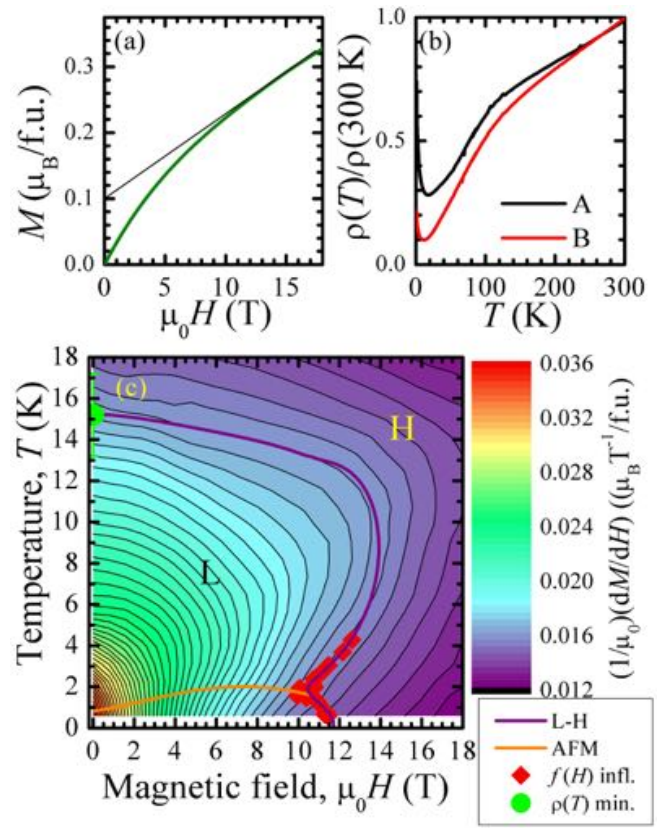}
\sloppypar
\caption{(a) Pulsed-field measurement of the
magnetization $M(H)$ of a CeOs$_4$Sb$_{12}$
crystal for ${\bf H} || {\bf c}$ and $T=0.58$~K.
The units of $M(H)$ are $\mu_{\rm B}$
per formula unit (f.u.).
Data are shown in dark green; the fine line
is shown as a guide to the eye, indicating the
rapidly increasing magnitude of ${\rm d}M/{\rm d}H$ as
the field falls below about 11~T.
(b)~Resistivity $\rho (T)$ normalized to the
value at $300$~K versus temperature for
two crystals (A, B) of CeOs$_4$Sb$_{12}$.
A minimum around $T\approx 15$~K
is clearly visible.
(c)~Phase diagram of 
CeOs$_4$Sb$_{12}$ including a 
contour plot of ${\rm d}M/{\rm d}H$
versus field $\mu_0H$ and temperature $T$.
The green point is the average position of
the minimum in resistance (see (b)) for
our samples, with the error bar showing the spread
of values; red points are the falling-field
inflection points from the PDO data
(see Fig.~\ref{CeSweep}).
The orange curve is the boundary of the
antiferromagnetic state, derived by averaging the
points of various authors shown in Fig.~\ref{CeSweep}(b),
but truncated by the L to H transition at high fields.
The purple curve is the notional phase boundary between the
 H and L phases.}
\label{CePhase}
\end{figure} 

Section~II described how the PDO response
is a convolution of the electrical resistance {\it and}
the (magnetic) differential susceptibility~\cite{saman};
in the phase-transition mechanism described above,
both of these quantities are expected to change at the L to H
boundary, with unpredictable consequences for the
PDO frequency~\cite{PDOfail}.
We therefore attempt a better delineation of the
L to H phase boundary using magnetometry.
Fig.~\ref{CePhase}(a) shows a pulsed-field measurement of
the magnetization $M(H)$ at a temperature of 0.58~K. 
A change in the behavior of $M(H)$ occurs at a field close 
to the shift in PDO frequency that we associate with
the L to H transition (Fig.~\ref{CeSweep}); at high fields 
the differential susceptibility ${\rm d}M/{\rm d}H$ is
relatively small, whereas at lower fields it is larger and
varies rapidly, leading to a curved $M(H)$ trace.
The fine black line in Fig.~\ref{CePhase}(a) is provided
as a guide to the eye in distinguishing the two regimes.
Note that the $\alpha$ (L) to $\gamma$ (H) transition in
Ce alloys is also accompanied by an analogous, fairly gradual, change in
${\rm d}M/{\rm d}H$~[\onlinecite{fivos}].

Free-energy considerations~\cite{dzero} show that increases
in temperature also favor quasi-localized $4f$ electrons;
therefore, increasing temperature should drive the system through the 
L to H transition at zero magnetic field.
Fig.~\ref{CePhase}(b) shows zero-field
temperature-dependent resistivity ($\rho(T)$)
data for two of the CeOs$_4$Sb$_{12}$ samples used in this
study; a distinct minimum is seen close to $T=15$~K.
The L phase, with its small number of heavy
quasiparticles, will have a high $\rho$ at low temperatures,
whereas the H phase, with its simpler, larger Fermi surface of
lighter quasiparticles, should exhibit relatively conventional
metallic $\rho(T)$ behavior at higher temperatures;
therefore, we associate the minimum in $\rho(T)$ with the
boundary between these two regimes. 

Fig.~\ref{CePhase}(c) is a compilation of the various data
in an attempt to identify the complete L to H phase boundary.
The color scale and contours are differential susceptibility
$({\rm d}M/{\rm d}H)$ data derived by differentiating $M(H)$ curves such as that
in Fig.~\ref{CePhase}(a); the whole plot is an interpolation
of data from 20~T pulsed-field sweeps at temperatures
of 0.58, 1.33, 2.54, 3.12, 3.64, 4.30, 5.00, 5.50, 6.00, 6.50,
7.00, 7.50, 8.50, 9.00, 10.00, 11.00, 13.50, 15.00, 16.00
and 18.00~K, plus constant-field temperature sweeps ($40-2$~K)
in the SQUID at fields of 0.1, 0.5, 1, 2, 3, 4, 5, 6 and 7~T.
The resistivity minimum (Fig.~\ref{CePhase}(b)) and the
features from the low-temperature PDO data (Fig.~\ref{CeSweep})
are superimposed on the plot,
along with the boundary around the 
antiferromagnetic phase (orange curve).

As discussed above, the L phase seems to be characterized
by a relatively high ${\rm d}M/{\rm d}H$, and the H phase
by lower, less field-dependent values.
At temperatures above 1.5~K, our notional boundary between L and H phases
is therefore based on the idea that some characteristic value
of the differential susceptibility ({\it i.e.} a contour in Fig.~\ref{CePhase}(c))
separates the two states~\cite{CeFoot}. 
To this end, note that
the resistive minimum and the kinks in the
PDO data above 1.5~K lie at very similar values 
of ${\rm d}M/{\rm d}H$.
We therefore trace a notional L to H phase boundary
that connects these points, roughly following the ${\rm d}M/{\rm d}H$
contours (Fig.~\ref{CePhase}(c); purple curve).

Sections IIIA and IIIC describe how, below 1.5~K and above 10~T, 
we expect the
L to H transition and the boundary of the
antiferromagnetic phase to coincide;
below 1.5~K, the notional phase boundary follows
the PDO data to reflect
the change in character of the L to H transition
observed in this region (see Fig.~\ref{CeSweep}). 

At first sight, by analogy with other correlated-electron
systems~\cite{heavyfermion,merino,Singleton}, 
it is tempting to simply ascribe the antiferromagnetic 
phase to a region of quantum fluctuations around the 
point at which the L to H transition plunges to $T = 0$. 
However, the manner in which the susceptibility 
contours, and hence the notional L to H phase boundary, 
curve back towards the origin as the temperature is 
cooled below 10~K is suggestive of some more 
complex interplay. Related behavior has been seen in 
the phase diagrams of reduced-dimensionality 
antiferromagnets~\cite{Pinaki,Kartsovnik}; 
but by contrast, in those cases, the effect 
occurred on the high$-T$, low$-H$ side of the phase 
boundary and was attributed to the effect of thermal 
fluctuations on the free energy of the system~\cite{Pinaki}. 
In the case of CeOs$_4$Sb$_{12}$, the backwards 
curvature instead occurs in the vicinity of the 
low-temperature antiferromagnetic state 
[orange phase boundary in Fig.~\ref{CePhase}(c)]. 
It is therefore possible that the effect 
is due here to antiferromagnetic fluctuations, 
rather than thermal fluctuations, that act to 
destabilise the L phase, pulling the L to H boundary 
to lower fields as temperature is reduced towards 
the antiferromagnetic transition. 
Once antiferromagnetic order is established, the 
free-energy landscape changes and it appears 
that the L phase becomes more stable as the temperature 
is lowered further, such that the L to H and 
antiferromagnetic phase boundaries coincide and 
move to higher fields. 
This interplay between the Fermi-surface reconstruction 
(the L to H transition) and the antiferromagnetism 
can arise because these two phases are inextricably linked. 
Not only are the energy scales of the two transitions 
similar in this part of the phase diagram, but also 
the antiferromagnetism is expected to arise from the 
formation of an SDW, a process which depends crucially 
on the Fermi surface and its fluctuations~\cite{Singleton}.

\begin{figure}[t]
\centering
\includegraphics[width=8.5cm]{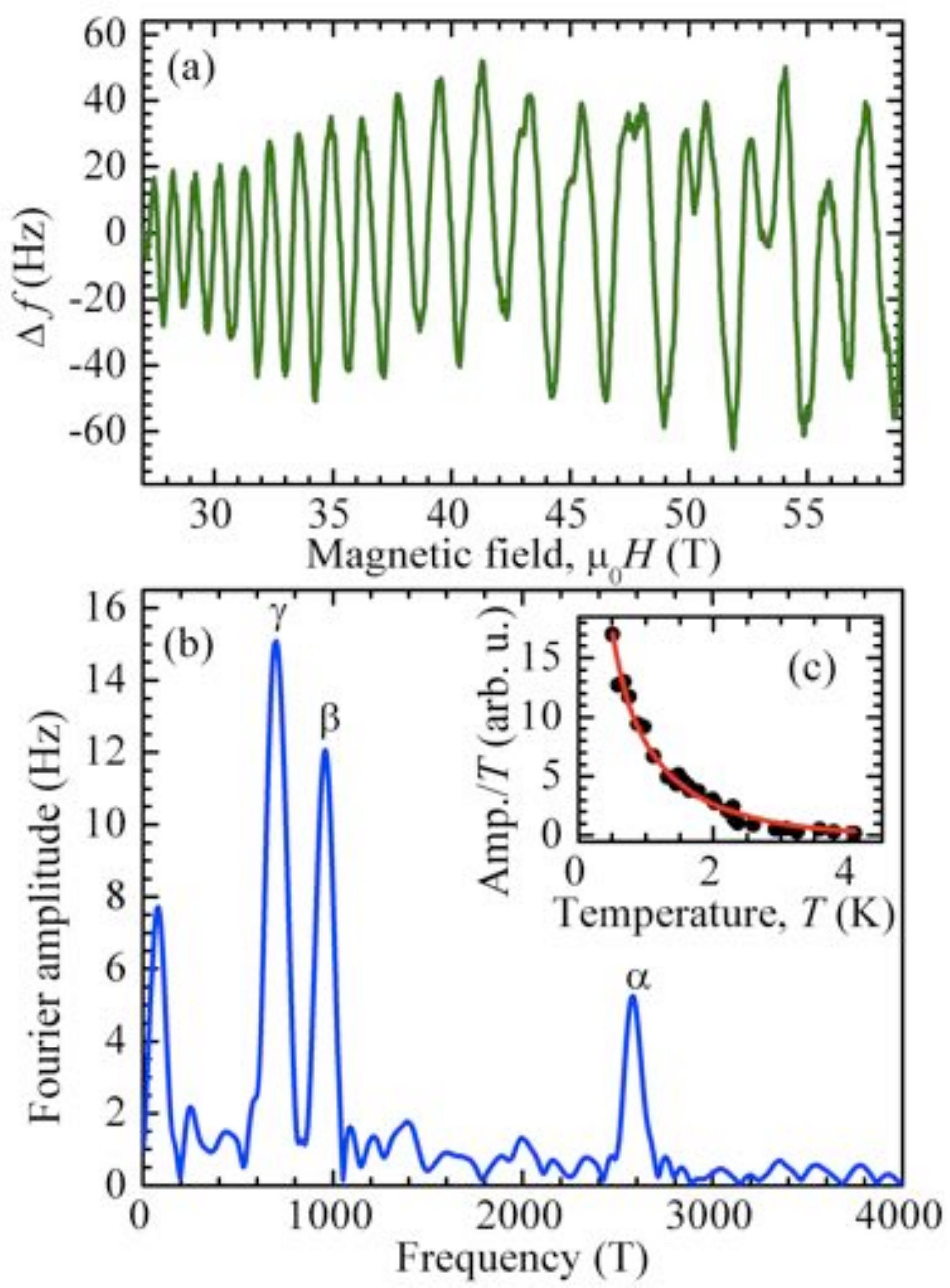}
\sloppypar
\caption{(a)~Shubnikov-de Haas oscillations 
in the PDO frequency shift of a NdOs$_4$Sb$_{12}$
single crystal,
obtained by subtracting the monotonic 
background dependence due to
magnetoresistance. The temperature is
$T=1.5$~K and the field is oriented at
$39.5^{\circ}$ to the $[0\bar{1}0]$ axis
for rotation about the $[110]$ axis.
(b) Fourier transforms of data similar to those
in (a) for {\bf H} making an angle of $6.7^{\circ}$
with respect to the $[0\bar{1}0]$ axis;
$T = 1.5$~K and the sample is
rotated about the $[110]$ axis.
The peak close to $90$~T is
thought to be an artefact of the background
subtraction routine.
The Fermi-surface sections are
labelled $\alpha$, $\beta$ and $\gamma$
following the scheme used
in PrOs$_4$Sb$_{12}$~[\onlinecite{Sugawara}].
(c)~NdOs$_4$Sb$_{12}$
$\alpha$-frequency Shubnikov-de Haas oscillation
Fourier amplitude divided by $T$ versus $T$;
data are points and the curve is a fit to
Equation~\ref{LK}.
{\bf H} was applied parallel to the [001] direction.}
\label{NdQOs}
\end{figure} 
In summary, the L to H transition in CeOs$_4$Sb$_{12}$
behaves in many ways ({\it i.e.} broadness, hysteresis,
gentle changes in resistivity and differential suceptibility)
very similarly to the $\alpha -\gamma$ (L-H) transition
observed in Ce and its alloys~\cite{fivos,dzero}.
The apparent change in Fermi surface from
the compensated semimetal predicted by bandstructure calculations
(L phase) to the single,
approximately spherical, much larger pocket
evidenced by the quantum oscillations in the H phase
is also in accord with such an explanation.
In effect, some of the charge carriers undergo a transition from
delocalized quasiparticles (L phase) to localized Ce 4f electrons
(H phase)~\cite{whine}.

\section{Neodymium Osmium Antimonide}
Figure~\ref{NdQOs}(a) shows typical quantum-oscillation data for
NdOs$_4$Sb$_{12}$; a Fourier transform
is given in Figure~\ref{NdQOs}(b).
Depending on the orientation of the
magnetic field, as many as five separate frequencies may be visible,
or as few as two. One of the frequencies
observed was very obviously a second harmonic,
suggestive of well-resolved spin splitting of the fundamental
frequency~\cite{shoenberg}.

\begin{figure}[t]
\centering
\includegraphics[width=8.5cm]{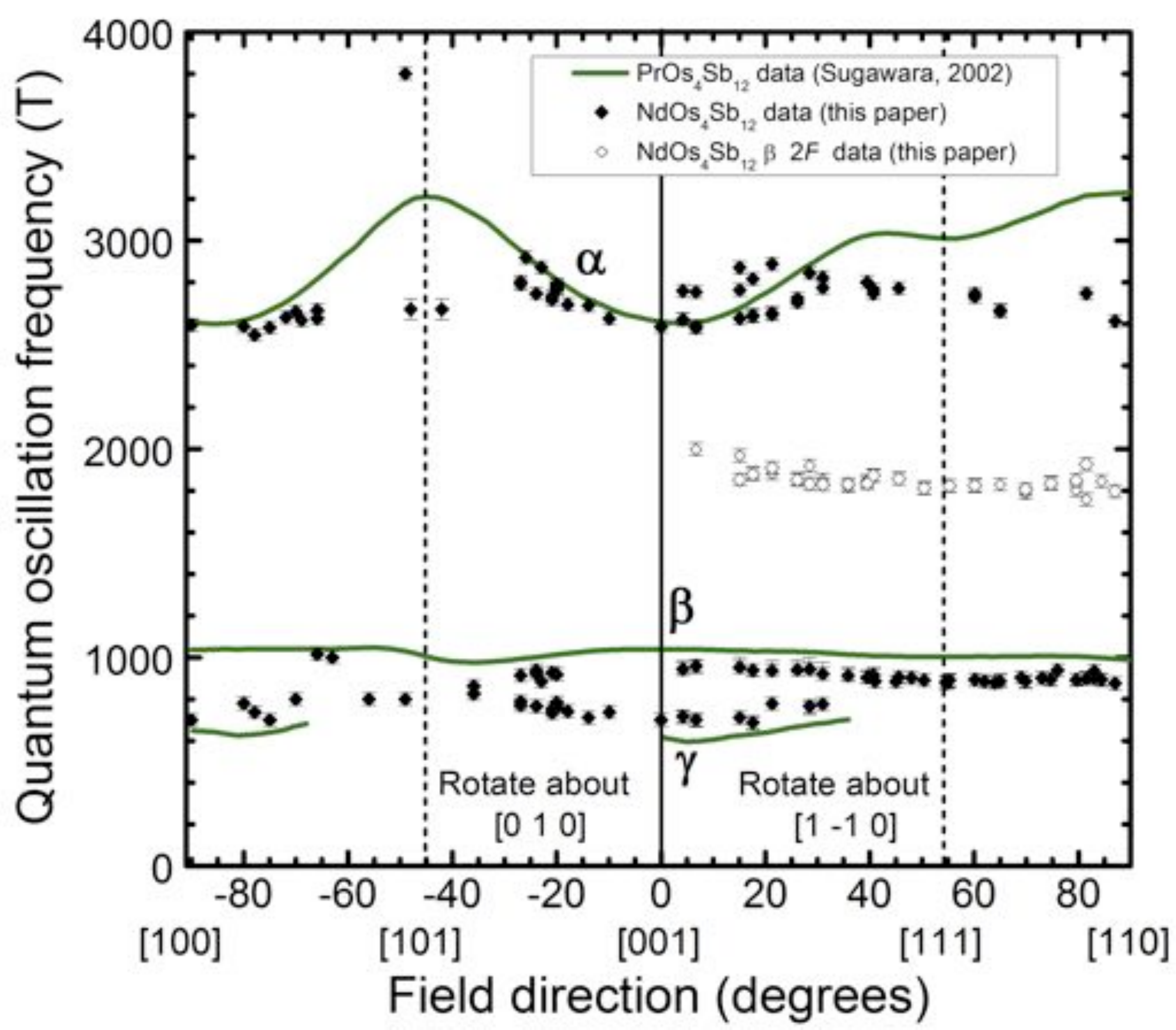}
\sloppypar
\caption{Quantum-oscillation frequencies versus
field orientation for NdOs$_4$Sb$_{12}$;
points correspond to well-resolved frequencies observed in the data;
the curves are analogous data from PrOs$_4$Sb$_{12}$
taken from Reference~\onlinecite{Sugawara}.
The Fermi-surface sections are labelled according to the
scheme used in the latter reference for PrOs$_4$Sb$_{12}$.
}
\label{freqplot}
\end{figure} 

\renewcommand{\arraystretch}{1.3}
\begin{table*}[t]
\centering
\begin{tabular}{|c|c|c|c|c|c|}
\hline
~ & ~ & 
\multicolumn{2}{c|} {~~~NdOs$_4$Sb$_{12}$ (this paper)~~~} & 
\multicolumn{2}{c|} {~~~PrOs$_4$Sb$_{12}$~(Ref.~\onlinecite{Sugawara})~~~} \\
\hline
Field orientation & Fermi-surface section & ~~~$F$ (T)~~~ & $m^*/m_{\rm e}$ & 
$F$ (T) & $m^*/m_{\rm e}$ \\ 
\hline 
${\bf H} ||$ [001] & $\alpha$ & $2560\pm 20$ & $3.1 \pm 0.2$  & 2610 & 4.1 \\
\hline
${\bf H} ||$ [001] & $\beta$ & $950\pm 20$ & $1.8 \pm 0.3$  & 1070 & 2.5 \\
\hline
${\bf H} ||$ [001] & $\gamma$ & $690\pm 30$ & $3.6\pm 0.4$ & 710 & 7.6 \\
\hline
${\bf H} ||$ [011] & $\beta$ & $870\pm 30$ & $1.5\pm 0.4$ & 875 & 3.9 \\
\hline
\end{tabular}
\caption{Effective masses and quantum-oscillation frequencies 
for NdOs$_4$Sb$_{12}$ and PrOs$_4$Sb$_{12}$.}
\label{effmass}
\end{table*}

The frequencies observed in NdOs$_4$Sb$_{12}$
are plotted in Figure~\ref{freqplot}
as a function of the orientation of the magnetic field.
For comparison, frequencies measured in 
de Haas-van Alphen experiments on PrOs$_4$Sb$_{12}$
reported in Ref.~\onlinecite{Sugawara}
are shown as continuous curves; for convenience,
we use the same cross-section labels as that work.
Overall, the Fermi-surface topologies of the two materials
are similar, with the $\beta$ Fermi-surface section
being slightly smaller in NdOs$_4$Sb$_{12}$ 
than in the Pr-compound. The $\alpha$ sheet
of the two materials is also similar, with that in
NdOs$_4$Sb$_{12}$ being somewhat slimmer for fields
close to the $[101]$ directions. 
By contrast, the $\gamma$ sheet is somewhat
larger in NdOs$_4$Sb$_{12}$.

A single, higher
frequency around 3800~T was observed for
fields close to alignment with $[101]$;
though an equivalent is not seen in experimental
data for PrOs$_4$Sb$_{12}$, it
is very similar to one of the frequencies predicted from
that material's bandstructure calculations~\cite{Sugawara}.
For this particular Fermi-surface section~\cite{Sugawara},
when ${\bf H} || [101]$,
it appears that the Fermi surface cross-sectional area
changes relatively slowly
as one moves away from the extremal orbit
along a direction ${\bf k} || {\bf H}$.
Such a topology will give an
enhanced quantum-oscillation amplitude
for a narrow range of angles around this
field orientation~\cite{shoenberg}. Moreover,
as discussed below, the effective masses are
generally lower in NdOs$_4$Sb$_{12}$
than in PrOs$_4$Sb$_{12}$,
favoring the observation of
quantum oscillations in the former.
  
Band structure calculations for PrOs$_4$Sb$_{12}$
predict a splitting of the $\alpha$ frequency
for fields around the [001] direction;
this was not observed in PrOs$_4$Sb$_{12}$~[\onlinecite{Sugawara}]
but seems to be resolved in the present experiments on
NdOs$_4$Sb$_{12}$.

Effective masses $m^*$ were derived from the temperature
dependence of the oscillation amplitudes
as shown in Figure~\ref{NdQOs}(c);
values are tabulated in Table~\ref{effmass},
and analogous data for PrOs$_4$Sb$_{12}$ are shown for comparison.
The masses for NdOs$_4$Sb$_{12}$
are consistently smaller than their equivalents in PrOs$_4$Sb$_{12}$,
suggesting that the interactions that lead to mass renormalization are
smaller in the former compound.

As mentioned in the Introduction, 
NdOs$_4$Sb$_{12}$ is a ferromagnet with an
ordering temperature $T_{\rm C} \approx 1$~K 
displaying mean-field ferromagnetism~\cite{Ho2005}. 
Its temperature-linear heat-capacity coefficient is
$\gamma\approx 520~{\rm mJmol^{-1}K^{-2}}$;
that for PrOs$_4$Sb$_{12}$ is 
$\gamma \approx 650~{\rm mJmol^{-1}K^{-2}}$~[\onlinecite{Sugawara}].
The fact that the effective masses are  significantly larger
in the latter than the former (Table~\ref{effmass}), whilst the
$\gamma$ values are relatively close,
suggests that another contribution to $\gamma$
might be present in NdOs$_2$Sb$_{12}$.
Theoretical models indicate that the off-center 
potential minima of the six rare earth sites 
(see Figure~\ref{fig1}(a)) could give 
rise to local phonons and enhance the electron-phonon 
interaction in the filled 
skutterudites.~\cite{Hattori2006,Hattori2005b, Hotta2007}.  
This effect results in Kondo-like phenomena, but with a non-magnetic 
origin; it can also contribute to a large electronic $\gamma$.  
These models have been used to explain the large magnetic-field-independent $\gamma \approx 820~{\rm mJmol^{-1}K^{-2}}$ in 
SmOs$_4$Sb$_{12}$.~\cite{Yuhasz2005,Sanada2005} 
Previous ultrasonic measurements show an extra mode in 
NdOs$_4$Sb$_{12}$ at
$T\approx 15$~K~[\onlinecite{Yanagisawa2007,Yanagisawa2008a,Yanagisawa2008b}];
it and the above-mentioned SmOs$_4$Sb$_{12}$ 
are the only two skutterudites displaying this 
ultrasonic anomaly~\cite{Yanagisawa2007,Yanagisawa2008a,Yanagisawa2008b}. 
The contrast between the relatively low cyclotron
effective masses in NdOs$_4$Sb$_{12}$
and its relatively high $\gamma$ value is therefore probably due to the
local phonon modes.

\section{Conclusions}
Magnetometry and high-frequency
conductivity measurements have been 
made on single crystals of CeOs$_4$Sb$_{12}$ 
and NdOs$_4$Sb$_{12}$ using temperatures down to 0.5~K 
and magnetic fields of up to 60~tesla.
The data show that CeOs$_4$Sb$_{12}$ undergoes a field-induced transition
from a semimetal L phase
into a high-field, high-temperature H phase 
characterized by a single, almost spherical
Fermi-surface section.
The general behaviour of the phase transition
and comparisons with models of the bandstructure lead us to
propose that the field-induced phase transition in
CeOs$_4$Sb$_{12}$ is similar in origin to the
well-known $\alpha-\gamma$ transition in Ce and
its alloys~\cite{fivos}.
The phase boundary appears to exhibit an unusual
curvature at low temperatures, possibly due to
quantum fluctuations associated with the
antiferromagnetic phase.
By contrast, the behavior of NdOs$_4$Sb$_{12}$
is relatively straightforward; it has
a similar Fermi surface topology to that of 
PrOs$_4$Sb$_{12}$ , but with significantly smaller effective masses. 
This result supports the importance of local phonon
modes in contributing to the low-temperature heat capacity of
NdOs$_4$Sb$_{12}$.
The original motivation for this study --
to investigate whether the superconductivity in PrOs$_4$Sb$_{12}$
arises from close proximity to AFM and FM quantum-critical points,
with the more explicitly magnetic
CeOs$_4$Sb$_{12}$ and NdOs$_4$Sb$_{12}$ 
lying on the either side of the postulated
quantum phase transitions -- is borne out in part.
NdOs$_4$Sb$_{12}$ and PrOs$_4$Sb$_{12}$ indeed look
like close relatives, with the heavier quasiparticle masses
in the latter favoring superconductivity.
But any resemblance of CeOs$_4$Sb$_{12}$
to the other two Skutterudites is overwhelmed by the
Kondo physics that is often writ large in Ce and
its compounds.

\section{Acknowledgments}
Research at CSU-Fresno is supported by NSF DMR-1506677. 
Work at UCSD is supported by
the US Department of Energy, Office of Basic Energy Sciences, 
Division of Materials Sciences and Engineering, under Grant 
No. DEFG02-04ER46105 (single crystal growth) and the 
National Science Foundation under Grant No. DMR~1206553 (sample characterization).
The portion of this work done at Hokkaido University is supported by 
JSPS KAKENHI grants nos. 26400342, 15K05882
and 15K21732.
Research at the University of Warwick is supported by EPSRC.
A portion of this work was performed at the National High
Magnetic Field Laboratory, which is supported by National
Science Foundation Cooperative Agreement No. DMR-
1157490, the State of Florida, and the U.S. Department of
Energy (DoE) and through the DoE Basic Energy Science
Field Work Proposal ``Science in 100 T''. 
JS thanks the University of
Oxford for provision of a visiting professorship that permitted
the low-field measurements featured in this paper.
We are very grateful to H. Harima for many illuminating comments
on an earlier version of this manuscript.
\vspace{5mm}

\noindent
$^*$Contact email pcho@csufresno.edu\\
$^\aleph$Contact email jsingle@lanl.gov\\
$^\dag$Current address: Johns Hopkins University Applied Physics
Laboratory, 11100~Johns Hopkins Road, Laurel MD~20723, U.S.A.\\
$^\ddag$Current address: Lawrence Livermore National Laboratory,
P.O.~Box~808, Livermore, CA~94551-0808.


\begin{thebibliography}{99}
\bibitem{tayama2015}
T. Tayama, W. Ohmachi, M. Wansawa, D. Yutani,
T. Sakakibara, H. Sugawara and H. Sato,
J. Phys. Soc. Jpn. {\bf 84}, 104701 (2015).
\bibitem{yan}
Binghai Yan, Lukas M\"{u}chler, Xiao-Liang Qi,
Shou-Cheng Zhang and Claudia Felser
Phys. Rev. B {\bf 85}, 165125 (2012).
\bibitem{HoNJP}
Pei-Chun Ho, J. Singleton, M.B. Maple, 
Hisatomo Harima, 
P.A. Goddard, Z. Henkie 
and A Pietraszko,
New J. Phys. {\bf 9}, 269 (2007).
\bibitem{aokireview}
Y. Aoki, H. Sugawara, H. Harima, and H. Sato, 
J. Phys. Soc. Jpn. {\bf 74}, 209 (2005).
\bibitem{oneprime}
M.B. Maple, E.D. Bauer, N.A. Frederick,
P.-C. Ho, W.A. Yuhasz, 
and V.S. Zapf, Physica B {\bf 328}, 28 (2003).
\bibitem{meisner}
G.P Meisner, M.S. Torikachvili, K.N. Yang, M.B. Maple 
and R.P. Guertin, J. Appl. Phys. {\bf 57}, 3073 (1985).
\bibitem{Sugawara1}
H. Sugawara, T.D. Matusda, K. Abe, Y. Aoki, H. Sato,
S. Nojiri, Y. Inada, R. Settai and Y. Onuki, 
Phys. Rev. B {\bf 66}, 134411 (2002).
\bibitem{Sugawara}
H. Sugawara, S. Osaki, S.R. Saba, Y. Aoki,
H. Sato, Y. Inada, H. Shishido, R. Settai, Y. Onuki,
H. Harima and K. Oikawa, Phys. Rev. B
{\bf 66}, 220504(R) (2002).
\bibitem{Sugawara3}
H. Sugawara, K. Abe, T.D. Matsuda, Y. Aoki,
H. Sato, R. Settai and Y. Onuki,
Physica B {\bf 312-313}, 264 (2002).
\bibitem{BauerED2001} 
E. D. Bauer, A. \'{S}lebarski, E. J. Freeman, 
C. Sirvent, and M. B. Maple, J. Phys.: 
Condens. Matter {\bf 13}, 4495 (2001).
\bibitem{Miyake2003}
K. Miyake, H. Kohno and H. Harima,
J. Phys.: Condens. Matter {\bf 15}, L275 (2003).
\bibitem{Ho2005}
P.-C. Ho, W. M. Yuhasz, N. P. Butch, N. A. Frederick, 
T. A. Sayles, J. R. Jeffries, M. B. Maple, J. B. Betts, 
A. H. Lacerda, P. Rogl, and G. Giester, 
Phys. Rev. B {\bf 72}, 094410
(2005).
\bibitem{heavyfermion}
John S. Van Dyke, Freeke Massee, Milan P. Allan, J.C. Seamus Davies,
Cedomir Petrovic and Dirk K. Morr, PNAS {\bf 111}, 11663 (2014), 
and references therein.
\bibitem{merino}
Jaime Merino and Ross H. McKenzie,
Phys. Rev. Lett. {\bf 87}, 237002 (2001).
\bibitem{maplepnas}
M.B. Maple, N.P. Butch, N.A. Frederick, P.-C. Ho,
J.R. Jeffries, T.A. Sayles, T. Yanagisawa,
W.M. Yuhasz, Songxue Chi, H.J. Kang, J.W. Lynn,
Pengcheng Dai, S.K. McCall, M.W. McElfresh,
M.J. Fluss, Z. Henkie and A. Pietraszko,
PNAS {\bf 103}, 6783 (2006).
\bibitem{yuhaszprb}
W.M. Yuhasz, N.P. Butch, T.A. Sayles, P.-C. Ho, J.R. Jeffries,
T. Yanagisawa, N.A. Frederick, M.B. Maple, Z. Henkie, A. Pietraszko,
S.K. McCall, M.W. Mc Elfresh and M.J. Fluss,
Phys. Rev. B {\bf 73}, 144409 (2006). 
\bibitem{saman}
S. Ghannadzadeh, M. Coak, I. Franke, P. A. Goddard, 
J. Singleton and J. L. Manson
Rev. Sci. Instrum. {\bf 82}, 113902 (2011).
\bibitem{moaz}
M.M. Altarawneh, C.H. Mielke and J.S. Brooks,
Rev Sci Instrum. {\bf 80}, 066104 (2009).
\bibitem{nhmfl}
J. Singleton, C.H. Mielke, A. Migliori,
G.S. Boebinger and A.H. Lacerda,
Physica B {\bf 346}, 614 (2004)
and references therein.
\bibitem{Goddard}
P. Goddard, J. Singleton, P. Sengupta, 
R. D. McDonald, T. Lancaster, S. J. Blundell, 
F. L. Pratt, S. Cox, N. Harrison, J. L. Manson, H. I. Southerland, 
and J. A. Schlueter, 
New J. Phys. {\bf 10}, 083025 (2008).
\bibitem{Rotundu}
C.Rotundu and B. Andraka, Phys. Rev. B {\bf 73}, 144429 (2006).
\bibitem{Namiki2003}
T. Namiki, Y. Aoki, H. Sugawara, and 
H. Sato,  Acta. Phys. Polonica B {\bf 34}, 1161 (2003).
\bibitem{fivos}
F. Drymiotis, J. Singleton, N. Harrison, L. Balicas, A. Bangura,
C.H. Mielke, Z. Fisk, A. Migliori, J.L. Smith and J.C. Lashley,
J.Phys.: Condens. Matter {\bf 17}, L77 (2005).
\bibitem{Sugawara2005}
H. Sugawara, S. Osaki, M. Kobayashi, T. Namiki, 
S. R. Saha, Y. Aoki, and H. Sato, Phys. Rev. B {\bf 71}, 125127
(2005).
\bibitem{Yogi2005}
M. Yogi, H. Kotegawa, G. Q. Zheng, 
Y. Kitaoka, S. Ohsaki,
H. Sugawara, and H. Sato, 
J. Phys. Soc. Jpn. {\bf74}, 1950
(2005).
\bibitem{YangCP2006}
C. Yang, Z. Zhou, 
H. Wang, J. Hu, K. Iwasa, H. Sugawara,
and H. Sato, Rare Metals {\bf 25}, 550 (2006)
\bibitem{YangCP2007}
C. P. Yang, H. Wang, J. F. Hu, K. Iwasa,  K. Kohgi, 
H. Sugawara, and H. Sato, 
J. Phys. Chem. C. {\bf 111}, 2391 (2007).
\bibitem{Harrison}
N. Harrison, S.E. Sebastian, C.H. Mielke, A. Paris, M.J. Gordon
C.A. Swenson, D.G. Rickel, M.D. Pacheco,
P.F. Ruminer, J.B. Schillig, J.R. Sims, A.H. Lacerda,
M.-T. Suzuki, H. Harima and T. Ebihara,
Phys. Rev. Lett. {\bf 99}, 056401 (2007).
\bibitem{shoenberg}
D. Shoenberg, {\it Magnetic oscillations in metals}, Cambridge University
Press (Cambridge, 1984).
\bibitem{HarimaPC}
H. Harima and K. Takegahara, J. Phys.: Condensed Matter
{\bf 15}, S2081 (2003).
\bibitem{dzero}
M.O. Dzero, L.P. Gor’kov and A.K. Zvezdin, J. Phys.: Condens. Matter
{\bf 12}, L711 (2000).
\bibitem{Sarrao1}
J.L. Sarrao, A.P. Ramirez, T.W. Darling, F. Freibert, A. Migliori,
C.D. Immer, Z. Fisk and Y. Uwatoko,
Phys. Rev. B {\bf 58}, 409 (1998).
\bibitem{Sarrao2}
C.D. Immer, J.L. Sarrao, Z. Fisk, A. Lacerda, C. Mielke and J.D. Thompson,
Phys. Rev. B {\bf 56}, 71 (1997).
\bibitem{Sarrao3}
J.L. Sarrao, Physica B {\bf 259-261}, 128 (1999).
\bibitem{Fi02}
D.C. Koskenmaki and K.A. Gschneider, p337, {\it Handbook on
the Physics and Chemistry of Rare Earths}, ed. 
K.A. Gschneider and L. Eyring (North Holland, Amsterdam,
1978).
\bibitem{Fi05}
J. Laegsgaard and A. Svane, Phys. Rev. B {\bf 59}, 3450 (1999).
\bibitem{Fi06}
K. Held, A.K. McMahan and R.T. Scaletter, Phys. Rev. Lett.
{\bf 87}, 276404 (2001).
\bibitem{beta}
The situation in pure Ce is complicated at ambient pressure
by the formation of a metastable dhcp $\beta$ phase;
this is conveniently avoided by studying Ce alloys
containing small fractions of La and Th
(see Ref.~\onlinecite{fivos} and references therein). 
\bibitem{CeFoot}
Similarly, no sharp feature in the differential susceptibility 
is seen in the case of the $\alpha$ to $\gamma$ transition
of Ce alloys~\cite{fivos}; instead, the phase boundary is
visible as a broad, rounded change in the gradient of $M(H)$ plots~\cite{fivos}.
\bibitem{HarimaAPS}
H.~Harima, private communication.
\bibitem{mybook}
J. Singleton, Chapter 8 in {\it Band Theory and Electronic Propeties of Solids}
(Oxford University Press, Oxford, 2002).
\bibitem{PDOfail}
As shown in Fig.~\ref{CeSweep}(b),
the upper boundary of the antiferromagnetic phase
is mostly rather flat, varying by less than 0.35~K between 3 and 10~T.
Our pulsed-field PDO measurements are constant-temperature
field sweeps, so that they are rather insensitive to this part
of the antiferromagnetic phase boundary;
we observe merely some slight changes in shape of the
$-\Delta f$ versus $H$ curves in this region of field and
temperature.
The situation is further complicated by the fact that
the PDO response is influenced both by changes in
suscpetibility {\it and} conductivity. 
\bibitem{Singleton}
J. Singleton, Rep. Prog. Phys. {\bf 63}, 1111 (2000).
\bibitem{Pinaki}
P. Sengupta, C.D. Batista, R.D. Mc Donald, S. Cox, J. Singleton, L. Huang,
T.P. Papageorgiou, O. Ignatchik,
T. Hermannsd\"{o}rfer, J.L. Manson, J.A. Schlueter,
K.A. Funk and J. Wosnitza, Phys. Rev. B {\bf 79},
060409 (2009).
\bibitem{Kartsovnik}
M. Kunz, W. Biberacher, N.D. Kushch, A. Miyazaki
and M.V. Kartsovnik, preprint arXiv1696.07331 (2016). 
\bibitem{whine}
We note
that the terms ``delocalized'' (corresponding roughly to
Cerium in the Ce$^{4+}$ state) and ``localized''
(corresponding to Ce$^{3+}$),
though frequently used to describe such
situations,
represent at best a qualitative handwave that somewhat obscures 
the true
quantum mechanics of the situation~\cite{dzero}. 
\bibitem{Imada2007}
S. Imada, H. Higashimichi, A. Yamasaki, 
M. Yano, T. Muro, A. Sekiyama, S. Suga, 
H. Sugawara, D. Kikuchi,
and H. Sato, 
Phys. Rev. B {\bf 76}, 153106 (2007). 
\bibitem{Hattori2006}
K. Hattori, Y. Hirayama, and K. Miyake, 
J. Phys. Soc. Jpn. {\bf 75} Suppl., 238 (2006).
\bibitem{Hattori2005b}
K. Hattori, Y. Hirayama, and K. Miyake, 
J. Phys. Soc. Jpn. {\bf 74}, 3306 (2005). 
\bibitem{Hotta2007}
T. Hotta, J. Phys. Soc. Jpn. {\bf 76}, 023705 (2007). 
\bibitem{Yuhasz2005}
W. Y. Yuhasz, N. A. Frederick,  P.-C. Ho, 
N. P. Butch, B. J. Taylor, T. A. Sayles, 
M. B. Maple, 
J. B. Betts, A. H. Lacerda, P. Rogl, 
and G. Giester, 
Phys. Rev. B {\bf 71}, 104402 (2005).
\bibitem{Sanada2005}
S. Sanada, Y. Aoki, H. Aoki, A. Tsuchiya, D. Kikuchi,
H. Sugawara and H. Sato, J. Phys. Soc. Jpn. {\bf 74}, 246 (2005).
\bibitem{Yanagisawa2007}
T. Yanagisawa, W. M. Yuhasz, P.-C. Ho, M. B. Maple, 
H. Watanabe, T. Ueno, Y. Nemoto, 
and T. Goto, J. M. M. M. {\bf 310}, 223 (2007).
\bibitem{Yanagisawa2008a}
T. Yanagisawa, W. M. Yuhasz, P.-C. Ho, 
M. B. Maple, H. Watanabe, Y. Yasumoto, Y. Nemoto, 
and T. Goto, Physica B {\bf 403}, 735 (2008).
\bibitem{Yanagisawa2008b}
T. Yanagisawa, P.-C. Ho, W. M. Yuhasz, 
M. B. Maple,
Y. Yasumoto, H. Watanabe, Y. Nemoto, 
and T. Goto, 
JSPJ {\bf 77}, 074607 (2008).
\end{thebibliography}
\end{document}